# Differentiable Programming for Inverse Estimation of Soil Permeability and Design of Duct Banks

Anusha Vajapeyajula[1] and Krishna Kumar[2]

[1]Maseeh Department of Civil, Architectural and Environmental Engineering, University of Texas at Austin, Texas, USA.

Email: anushav@utexas.edu

[2]Assistant Professor, Maseeh Department of Civil, Architectural and Environmental Engineering, University of Texas at Austin, Texas.

Email: krishnak@utexas.edu

**ABSTRACT**

Underground duct banks carrying power cables dissipate heat to the surrounding soil. The amount of heat dissipated determines the current rating of cables, which in turn affects the sizing of the cables. The dissipation of heat through the surrounding soils happens through conduction and convection. The mode of heat transfer depends on the soil's thermal and hydraulic properties like diffusivity and permeability. The soil surrounding the cables could be designed to have maximum heat dissipation to have an improved current rating of cables. Differentiable programming is a novel technique that combines automatic differentiation with gradient-based optimization to minimize a loss function. Hence, differentiable programming can be used to evaluate input parameters based on output results. Given a desired heat distribution in the soil and a temperature source, we use differentiable programming to solve the inverse problem of estimating the soil permeability. In the present study, we employ differentiable programming to optimize the design of the buried duct bank and the backfill soil to improve heat dissipation. The design involves optimizing the permeability and size of the fill material compared to the surrounding natural soil. To implement automatic differentiation, we develop an inverse finite difference code in the Julia programming language and ForwardDiff package. We demonstrate the design capabilities of the differentiable programming technique to obtain the optimum permeability of the backfill material from the norm of the temperature distribution in the surrounding soil.

**INTRODUCTION**

Duct banks, housing power cables, are subjected to soil-mediated heat dissipation dynamics. The cable's current rating, crucial for its sizing, hinges on the inherent thermodynamics. Heat originating from the cables dissipates through the adjacent soil and finally into the air. Soil's thermal and hydraulic properties direct the heat dissipation mechanism, predominantly via conduction and convection. An engineered fill, encapsulating the cable, can be designed with tailored hydraulic properties to maintain optimal temperatures for proficient heat dissipation.

Heat transfer from underground duct banks has traditionally been assumed to be through the process of conduction (Neher and McGrath 1994; IEC 60287-1-1 2006). Conduction is the transfer of heat through thermal resistance. The heat dissipated through conduction is uniform all around the source of heat. In saturated soils, the mechanism of convection, where the heat transfer is due to the density difference created by temperature, also plays an important role in the heat dissipation. This results in an overall heat dissipation directed upwards (Kumar et. al, 2021).

The current rating of power cables can be improved with the right level of heat dissipation to the surrounding soil. The dependence of the heat transfer on permeability of the soil could indirectly help in assessing the optimal permeability of soils for maintaining the required temperature distribution and thus help in designing the backfill. The optimum permeability in a finite difference model of the soil can be assessed using a tool called Differentiable programming.

In Differentiable Programming (DP), the derivatives of functions are automatically computed in a high-level programming language. DP allows for gradient-based optimization of parameters to approximate a loss function using the concepts of gradient-descent and automatic differentiation. Automatic differentiation (AD) uses symbolic rules of differentiation to provide numerical values of derivatives (Baydin et. al, 2018) and therefore has a two-sided nature - partly symbolic and partly numerical (Griewank, 2003). Utilizing AD, both the function and its derivative are computed simultaneously per computational step, enhancing the efficiency of solving inverse problems where initial parameters are inferred through the gradient descent calculation of loss function.

In this paper, we demonstrate the ability of differentiable programming in solving the inverse problem to compute the permeability by minimizing the norm of temperature distribution. We then design a backfill above the duct bank to maximize heat loss as a two layer system with natural soil permeability and backfill permeability. Finally, we obtain the optimum permeability of the backfill soil for a given final temperature distribution.

**METHODOLGY**

*Automatic differentiation:*

Automatic differentiation is based on the usage of Dual numbers while calculating the expression for a differential using Taylor's series. Dual numbers are numbers of the form a + bϵ where $\epsilon^2 = 0$. The Taylor series expansion of a Dual number is given as

$$f(a + \epsilon) = f(a) + \frac{f'(a)}{1!}\epsilon + \frac{f''(a)}{2!}\epsilon^2 + \cdots$$

Here, all the terms with powers of ϵ equal and above 2 will be zero as $\epsilon^2 = 0$ and the above expression reduces to

$$f(a + \epsilon) = f(a) + \epsilon f'(a)$$

So, we obtain the function and its derivative when we evaluate the function at a dual number. The solution is also exact because no higher-order terms are neglected like in numerical differentiation. As the number of terms involved in the computation are fewer, the computation speed would also be high.

In the computer implementation of Automatic differentiation, the programming language Julia uses the Forward mode of AD in its ForwardDiff package. The independent variable with respect to which the differentiation is performed is fixed in the forward mode. The derivative of each sub-expression is performed recursively.

Firstly, the function is represented as a computational graph called the Wengert list (Wengert, 1964) where each node represents the intermediate result of the computation. Suppose y is a composite function: y = f(g(h(x))), with y: $R^m \to R^n$, x $\in R^n$, f: $R^n \to R^k$; n<k, g: $R^k \to R^l$ and h: $R^l \to R^m$.

y = f(g(h(x))) = f(g(h($w_0$))) = f(g($w_1$)) = f($w_2$) = $w_3$ where, $w_i$ are the nodes of the Wengert list.

$w_0$ = x, $w_1$ = h($w_0$), $w_2$ = g($w_1$), $w_3$ = f($w_2$) = y

$$\frac{\partial y}{\partial x} = \frac{\partial w_3}{\partial w_2} * \frac{\partial w_2}{\partial w_1} * \frac{\partial w_1}{\partial x}$$

As a general rule, $\frac{\partial w_i}{\partial x} = \frac{\partial w_i}{\partial w_{i-1}} * \frac{\partial w_{i-1}}{\partial x}$ is recursively computed in the forward mode. At each node, the value of the function and its derivative are computed and stored and finally, the chain rule is used to combine the intermediate results to obtain the final result. The JIT compilation of Julia (Bezanson et. al, 2014) is used by ForwardDiff to transparently recompile user code, to effectively implement higher order differentiation (Revels et. al, 2016).

We use Julia with the ForwardDiff package in solving the heat transfer problem.

*The forward problem:*

The first step in the process is the forward problem where we obtain the temperature distribution across the soil volume resulting from a known initial temperature distribution, soil thermal properties and permeability. We use a 2D Finite Difference model to solve the forward problem. We consider both conductive and convective heat transfer mechanisms by solving the partial differential equations for the heat transfer and the time-independent coupled fluid flow.

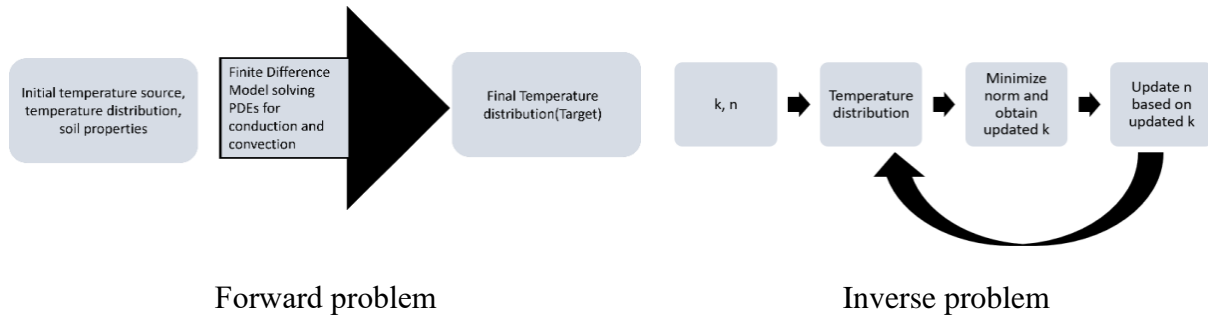

Forward problem            Inverse problem

**Figure 1: Forward and Inverse Problems**

*Governing equations:*

The heat transfer in the presence of a constant source under transient state is given by:

dT/dt = -α $\nabla^2 T$ + u $\nabla T$            *Eq. 1*

where α is the thermal diffusivity of the soil ($m^2$/s), $T$ is Temperature (K) and u is the fluid velocity (m/s).

It is assumed that the soil pores are fully saturated with water and therefore the properties of the soil-water medium are calculated using the porosity(n) of soil using the Eq. 2 and Eq. 3 below:

Thermal conductivity of the medium ($\lambda$):

$$\lambda = \lambda_s * (1-n) + \lambda_f * n \qquad Eq.\ 2$$

Specific heat of the medium (Cp):

$$C_p = C_{ps} *(1-n) + C_{pf} * n \qquad Eq.\ 3$$

where $\lambda_s$ and $\lambda_f$ are the thermal conductivities (W/m - K) of the soil particles and the fluid in the pore spaces respectively and $C_{ps}$ and $C_{pf}$ are the specific heat capacities (J/kg- K) of soil particles and pore fluid respectively.

In a fully saturated soil, Darcy's law describes the fluid flow through the porous media resulting from density gradient $\rho = \rho_0(1-\beta\Delta T)$, caused by the thermal differences as

$$u = \frac{-1}{n\mu} k * (\nabla p + g\, \rho_{f0}\, (1 - \beta(T - T_0))) \qquad Eq.\ 4$$

where $u$ is the fluid flow velocity (m/s), $n$ is the porosity, $\mu$ dynamic viscosity (Pa- s), $k$ is the intrinsic permeability (m2), $p$ is pressure (Pa), $\rho_{f0}$ is the reference fluid density at ambient temperature, and $\beta$ is the volumetric coefficient of thermal expansion (1/K).

As there is no pressure gradient, Eq. 4 reduces to

$$u = \frac{-1}{n\mu} k * g\, \rho_{f0}\, (1 - \beta(T - T_0)) \qquad Eq.\ 5$$

We solve the forward problem using central difference in space and an upwind scheme in time to model the temperature field (T).

### *The Inverse Problem:*

The inverse problem requires computing the permeability of the soil given a heat distribution. In this process, the temperature distribution for an estimate of permeability ($k_f$ * target permeability) is calculated and the loss is calculated as the norm of the difference in the obtained and target temperature distributions. Here $k_f$ is the factor by which the permeability is multiplied and is called permeability factor. The loss computed is used in the back calculation of permeability which minimizes the loss. New updates of permeability are obtained using the Newton Raphson method as shown in Eq. 6 and for each update of permeability, the porosity is also updated using the Newton Raphson method as shown in Eq. 7.

$$k = k - \frac{f}{df/dk} \qquad Eq.\ 6$$

Where f is the norm of the difference of temperature distribution

$$n = n - \frac{fn}{dfn/dn} \qquad Eq.\ 7$$

where fn is the difference in the updated and original permeabilities.

While using the above equations in computations, Automatic Differentiation is useful as both the function and its derivative are computed together in every step of the inverse problem.

**Material Properties:**

Table 1 presents the properties of the soil fill and thermal properties of soil and fluid used in the study. As the soil is silty, the porosity has been assumed to be in the range of 0.4-0.45 and the permeability of the order of 1E-13 m².

### Table 1: Properties of the soil fill

| Properties | Values | Properties | Values |
|---|---|---|---|
| Porosity($n$) | 0.4 | Specific heat of soil ($C_{ps}$) | 800 J/kg-K |
| Grain size($d_m$) | 0.025 mm | Density of fluid($\rho_f$) | 1000 kg/m³ |
| Specific gravity ($Gs$) | 2.7 | Specific heat of fluid ($C_{pf}$) | 4180 J/kg-K |
| Density($\rho$) | 1850 kg/m³ | Viscosity of fluid($\mu$) | 0.001 kg/m-s |
| Thermal conductivity of soil ($\lambda s$) | 1.0 W/m-K | Thermal conductivity of fluid ($\lambda_f$) | 0.6 W/m-K |

**Mesh Properties:**

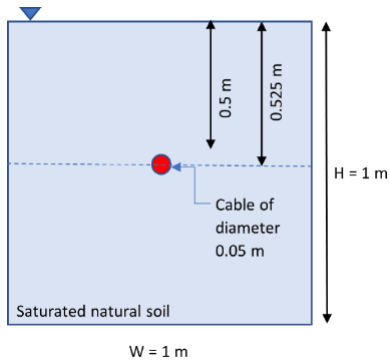

**Figure 2: Schematic of the simulation geometry (not to scale). A mesh with 40,000 elements is modeled.**

The temperature distribution of a 1m*1m grid comprising of elements of size dx = 0.005 m, dy = 0.005 m around a cable of radius 0.025 m is considered in the current model. We obtain the permeability of the soil grid using a mean particle size of $d_m$ = 0.0025 mm and a porosity of 0.4. The initial temperature of the cable is 30°C above the ambient temperature. Hence the temperature of the surrounding soil is taken as 0°C. Time step dt = 0.5s. We run forward and inverse problems for 100,000-time steps.

**RESULTS AND DISCUSSION:**

We obtain the heat transfer from a single cable which is at a temperature of 30°C above ambient temperature buried at a depth of 0.5 m inside a soil fill of size 1mx1m.

Using the semi empirical Karmen-Cozney equation, the target permeability (6.1728E-13 m$^2$) is computed using the target porosity value of 0.4 and mean particle size($d_m$). In the forward problem, the temperature distribution is found on the finite difference grid and the norm of the target temperature distribution is computed to be 561.31. We initialize the inverse problem with a porosity of 0.45 and a permeability factor of 0.005. This results in a temperature distribution with a norm of 507.39. The permeability in each consecutive iteration is updated based on the norm of the difference of target temperature and the calculated temperature distribution. The porosity is updated for the updated permeability value. With each iteration, the permeability and porosity approach the target values as the norm of temperature reaches closer to the norm of target temperature distribution. The tolerance for error in permeability is 1E-10 m$^2$ and in porosity is 1E-15.

Figure 3a shows the target temperature distribution for the forward problem. Figure 3b shows the evolution of loss function with each iteration. The variation of permeability and norm of temperature can be seen in Figure 3c.

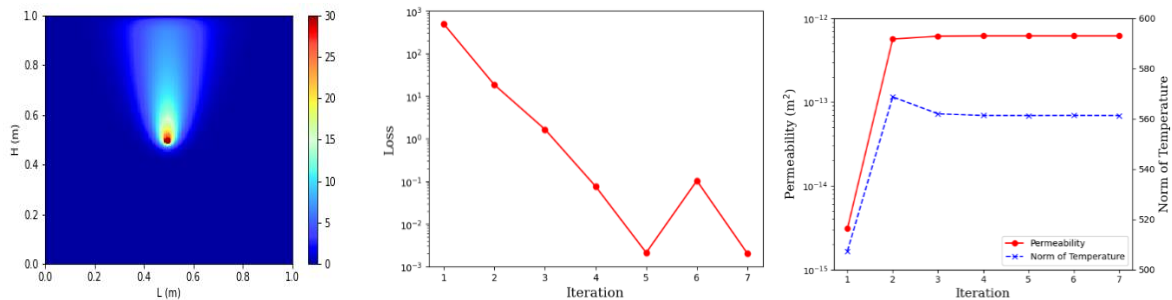

**Figure 3: a. Temperature distribution for the target values:
Permeability = 6.1728E-13 m$^2$, Target porosity = 0.4.
b. Plot showing the variation of loss with each iteration.
c. Plot showing variation in Permeability and Norm of Temperature with each iteration.**

Figure 4 shows the variation in the temperature distribution in the soil around the cable with iterations. We can see that the distribution inches closer to the target distribution with each iteration. The absolute error decreases by an order of magnitude with each iteration. By iteration 5, the predicted permeability is very close to the target permeability, demonstrating a quick convergence of the algorithm to the target value.

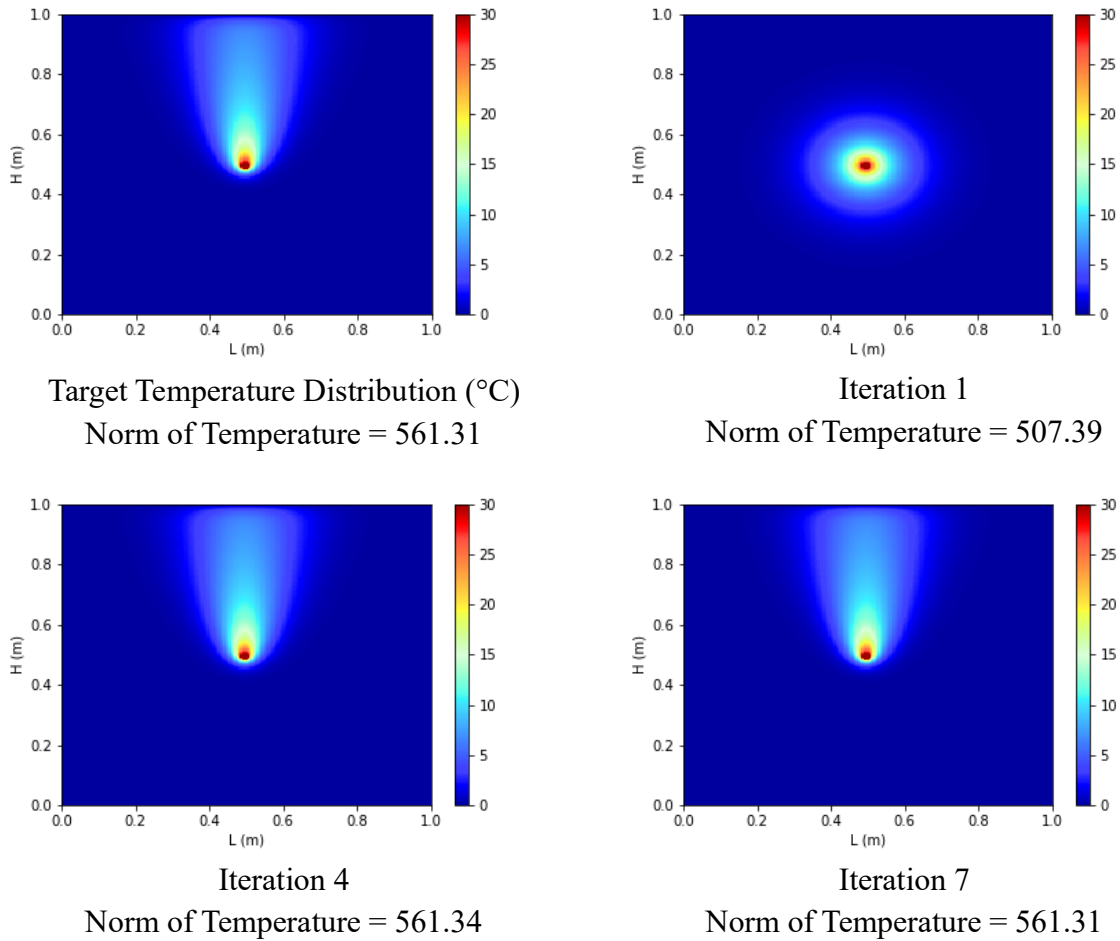

**Figure 4:  Variation in the temperature distribution in the soil fill with each iteration.**

The maximum error in the final iteration is of the order 7E-5 and the predicted permeability is 6.1727E-13 m$^2$ as compared to the target value of 6.1728E-13 m$^2$. This shows that the method of differentiable programming can be successfully used to predict the permeability by minimizing the loss in the norm of temperature distribution.

**Design:**
Now that the method of differentiable programming is validated, we design a soil fill with 2 types of soil. An inner layer of 0.5 m x 0.5 m containing soil with a permeability of 6.1728E-13 m$^2$ and the outer layer of clay with a permeability of 1E-16 m$^2$. The properties of the two layers are shown in Table 2.

**Table 2: Properties of the 2 soil layers in the design problem**

| Property | Fill | Natural soil (Clay) |
| --- | --- | --- |
| Thermal Conductivity of solids ($\lambda_s$) W/m-K | 1 | 1.5 |
| Specific heat capacity of solids ($C_{ps}$) J/kg-K | 800 | 2000 |
| Porosity | 0.4 | 0.6 |
| Density (kg/m$^3$) | 1900 | 1850 |
| d50 (mm) | 0.025 | 0.002 |
| Thermal diffusivity (α) m$^2$/s | 2.054E-07 | 1.57E-07 |
| Permeability (m$^2$) | 6.17E-13 | 1.00E-16 |

A soil fill with 2 layers has been designed for 2 cables of diameter 0.05m placed two diameters apart about the center of the fill soil along the width. The fill consists of an inner soil with a higher permeability and outer natural soil as shown in Figure 5. The goal in this case is to predict the permeability of the inner soil type using differentiable programming. The target permeability is 6.1728E-13 m2 for a target porosity of 0.4. The norm of target temperature distribution is computed to be 827.937.

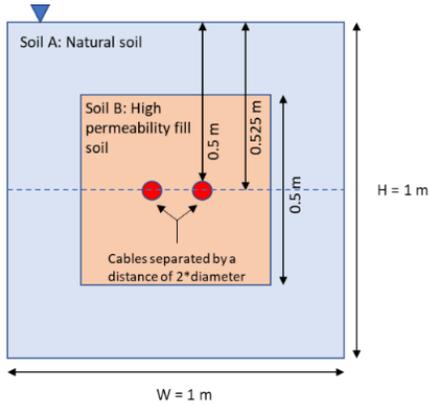

**Figure 5: Schematic for the design fill with 2 layers.**

As the loss decreases, the prediction gets closer to the actual value. This evolution of loss and the variation in permeability and norm of temperature can be seen in Figure 6b and Figure 6c respectively.

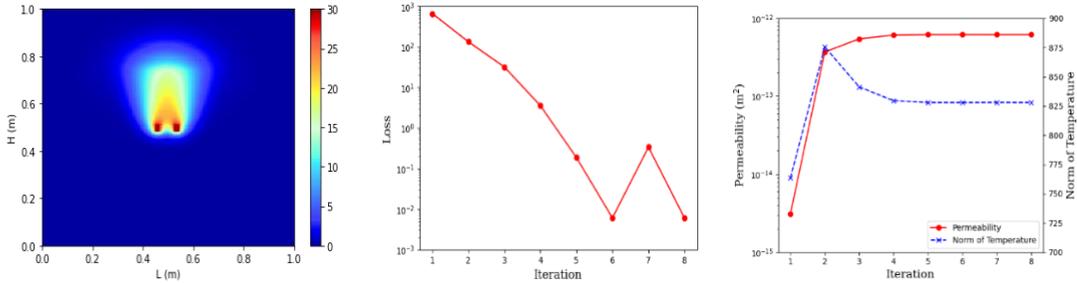

**Figure 6: a. Target Temperature distribution for the design fill**
**b. Plot showing the variation of loss with each iteration.**
**c. Plot showing variation in Permeability and Norm of Temperature with each iteration.**

The maximum error in the final iteration is of the order 1E-4 and the predicted permeability is 6.1726E-13 m$^2$ as compared to the target value of 6.1728E-13 m$^2$. This shows that the prediction in the design case is also very accurate. Figure 7 shows the variation in the temperature distribution in the soil around the cables for each iteration. We can see that by Iteration 8, the distribution very closely resembles the actual target temperature distribution.

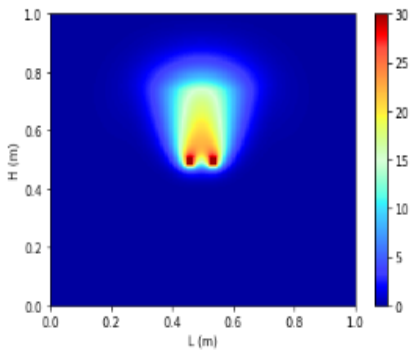

Target Temperature Distribution (°C)
Norm of Temperature = 827.94

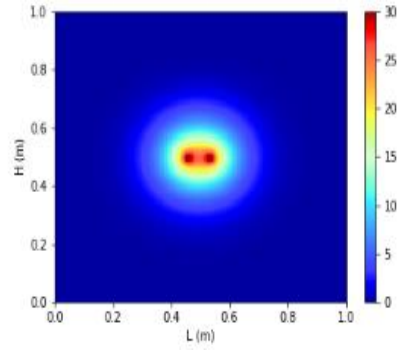

Iteration 1
Norm of Temperature = 763.42

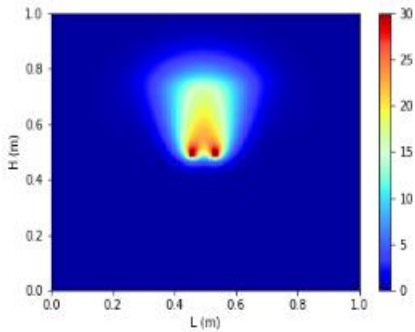

Iteration 3
Norm of Temperature = 841.33

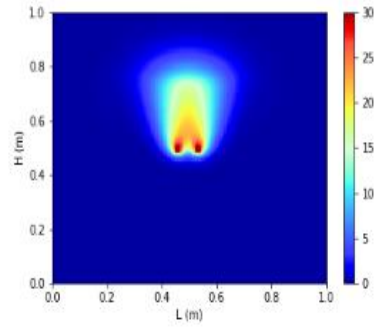

Iteration 5
Norm of Temperature = 828.02

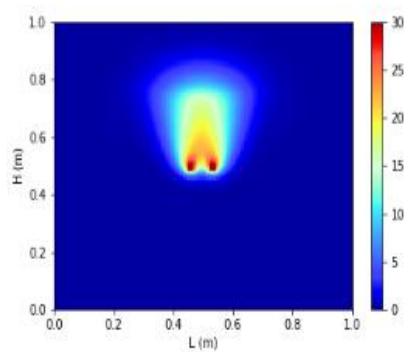

Iteration 8
Norm of Temperature = 827.94

**Figure 7: Variation in the temperature distribution in the soil fill with each iteration for the design problem**

# CONCLUSIONS

This study showcases the use of differentiable programming to solve the inverse problem of computing permeability by minimizing temperature distribution norms. It presents a two-layer backfill design surrounding a duct bank to optimize heat loss, considering both natural soil and backfill permeability. Consequently, we determine the optimal permeability of backfill soil for a specified final temperature distribution. While the current study involved the use of uniform soil parameters, the problem might become more complex in the case of layered natural soils with different properties.

Differentiable programming has considerable advantages in both prediction accuracy and computational efficiency, making it a promising tool for addressing complex inverse problems in various applications. Further work is being done to incorporate data from an experimental setup of a buried cable in soil to obtain the permeability of the soil (measured in the experiment) computationally using differentiable programming. This work paves the way for further advancements in optimization and inverse problem-solving for problems involving PDEs in Geotechnical Engineering.